\documentclass[preprint]{revtex4-1}

\usepackage{amsmath}
\usepackage{graphicx}
\graphicspath{ {images/} }

\DeclareMathOperator{\Tr}{Tr}

\newcommand{\kett}[1]{| #1 \rangle \rangle}
\newcommand{\bbra}[1]{\langle \langle #1 |}
\newcommand{\bbrakett}[2]{\langle \langle #1 | #2 \rangle \rangle}

\newcommand{\ket}[1]{| #1 \rangle }
\newcommand{\bra}[1]{\langle  #1 |}
\newcommand{\braket}[2]{\langle #1 | #2 \rangle}

\begin{document}

\title{Sequential measurements: Busch-Gleason theorem and its applications}
\author{Kieran Flatt }
\email{k.flatt.1@research.gla.ac.uk}
\author{Stephen M. Barnett}
\author{Sarah Croke}
\affiliation{School of Physics and Astronomy, University of Glasgow, Kelvin Building, University Avenue, Glasgow, G12 8QQ, UK}

\begin{abstract}
Probabilities enter quantum mechanics via Born's rule, the uniqueness of which was proven by Gleason. Busch subsequently relaxed the assumptions of this proof, expanding its domain of applicability in the process. Extending this work to sequential measurement processes is the aim of this paper. Given only a simple set of postulates, a probability measure is derived utilising the concept of Liouville space and the most general permissible quantum channel arises in the same manner. Super-Liouville space is constructed and a Bayesian interpretation of this object is provided. An important application of the new method is demonstrated, providing an axiomatic derivation of important results of the BB84 protocol in quantum cryptography.
\end{abstract}

\maketitle 

\section{Introduction}

The requirements for quantum security become ever more stringent as cryptographic and communicative routines are practically realised \cite{experimentalbb84, norbertreview}. A wide range of eavesdropping strategies are known, but in order that the two or more parties involved can feel safe a rigorous approach is needed. It is reassuring to know, in particular, precisely those features of quantum theory upon which the security relies. At the heart of this problem lies the topic of sequential measurements, with that performed by the intended recipient preceded by that of an eavesdropper. As an active process, so that the state is changed by measurement, any unauthorised access will leave a trace upon what is being sent. Disturbances of this kind are typically handled with the Kraus formalism \cite{kraus} but it is conceivable that this may not a priori cover all possibilities. Though the most general permissible maps are known, proofs rely on a presumed causal structure: a preparation event followed by one measurement (after which post-selection may occur) and subsequently another. It is known that this picture does not cover every aspect of the problem, and indeed recent research investigates correlations with indefinite causal structure \cite{causalorder}. Retrodictive formulations of quantum mechanics \cite{retrodiction1, retrodiction2} are one example and eavesdropping may be thought of as {\em interdictive}: information about two events is extracted from an interstitial measurement, i.e. is conditioned on both earlier and later measurements. It is also not obvious how delayed choice experiments would fit into such a prescription. For complete generality, and to be certain that no possibility has been overlooked, an axiomatic approach is desirable. Here it is assumed only that measurement outcomes may be associated with positive operators and that probabilities are linear in these. From these axioms it is shown that the most general structure of sequential measurements may be derived.
\par
Perhaps the single most important motivation for re-examining the formal structure of measurements comes from quantum key distribution, where the security of the channel must be established against the most general allowed eavesdropping strategy. In rederiving the results of the established Kraus formulation, more than established results are arrived at. It is determined precisely those features of quantum theory upon which the security of quantum key distribution relies.
\par
The seed for this work is a well-known theorem due to Gleason \cite{gleason}, essentially a statement that Born's rule is the unique way by which probabilities are calculated in quantum theory. It is proven that the only measure associated with a von Neumann measurement, given that it acts in a Hilbert space of two or more dimensions, can be given by the taking the trace over the relevant projector in product with another operator, the density matrix, which may be freely chosen. This result received much attention from Bell \cite{bellgleason} and others for the implications on theories of hidden variables, another area impacted by this work \cite{peres}. Much later on, the range of validity was extended to include qubits, generalised measurements (represented by POVMs as opposed to projectors) and the possibility of post-selection \cite{cavesgleason, busch, barnettetal, hall}. These generalised results are henceforth grouped together under the single title `Busch-Gleason theorem'.
\par
Relevant to the topic under discussion here are {\em conditional} probability measures specifically. Cassinelli and Zanghi \cite{cassinelli+zanghi} were the first to provide a Gleason-like theorem for two or more sequential measurements. That work is limited, however, as it considers only von Neumann measurements and also in that it privileges the non-general causal structure discussed above. In what follows, an analogous method is used which avoids both of these limitations. In essence, the approach is to exploit the properties of inner products, which can be naturally associated both with trace operations (in a mathematical sense) and probabilities (from physical arguments, and in keeping with previous work on axiomatic quantum mechanics \cite{hardy, infoaxioms}). Considering different spaces leads to both single and sequential measurement rules to be fixed by the requirement of positivity. The term measurement is used in a rather general sense and this may refer, for example, to an intermediate measurement including the state update, in between pre- and post-selection. In this way a characterisation of complete positivity is arrived at naturally and axiomatically. Further, this work shows how to condition on prior information, analogous with Bayes' rule for classical probability theory. 
\par
A related piece of work has recently been published by Shrapnel et al \cite{shrapneletal}, in which the Born and state-update rules are subsumed into a single postulate by taking completely positive maps as the foundational object of quantum mechanics. In what follows, positive operators on Hilbert space are the starting point and completely positive maps arise naturally in considering physical scenarios with pre- and post-selection. Shrapnel et al look to reinterpret certain objects in quantum theory (into the language of events and processes) while the aim here is to develop the logical structure from minimal mathematical assumptions. 
\par
Some background material is first introduced in order to clarify what is meant by the probability rules discussed above, especially in the context of sequential measurements. The Busch-Gleason theorem is then re-derived with this new toolkit, showing the result that probabilities are understood as inner products. Centrally, the extension to sequential measurements is performed by moving to a higher dimensional space which recontexualises the role of the three events (i.e. the preparation, first and second measurement). The use of this new construction is demonstrated with analysis of the delayed choice quantum eraser experiment and BB84 protocol, giving a new take on some known results.

\section{Sequential measurements}

\begin{figure}
 \includegraphics[width=0.5\textwidth]{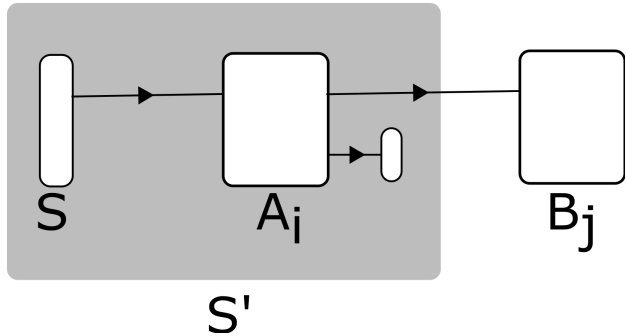}
 \caption{
 A measurement procedure can be visualised using this flowchart. A preparation procedure, here labelled $S$, will output a density matrix $\hat{\rho}$. The first measurement is represented by a set of Kraus operators $\{ \hat{A}_i \}$; outcome $A_0$ will occur with probability $\Tr(\hat{\rho} \hat{A}_0^{\dagger} \hat{A}_0 ) $ and the output state will be $\hat{A}_0 \hat{\rho} \hat{A}_0^{\dagger}$. The smaller box symbolises the possibility of post-selection and the shaded area labelled $S^{\prime}$ shows that the total procedure up to this point may be considered as a single preparation procedure. Finally, a second measurement occurs associated with Kraus operators $\{ \hat{B}_j \}$.} \label{fig:measurements}
  
 \end{figure}

There are two aspects to the act of measurement, as shown schematically in Fig. \ref{fig:measurements}. First, the probability distribution of outcomes is represented by a positive-operator valued measure (POVM) \textemdash a set of elements $\{ \hat{\pi}_i \}$ constrained by the requirements of Hermiticity, positivity and completeness. Second, there is a state change associated with a given result. In general, state changes are described by \textit{quantum channels}, input-independent maps between possible states. For the act of measurement, the most general permissible channel is \cite{kraus}
\begin{equation} \label{eq:choi}
\hat{\rho} \rightarrow \hat{\rho}^{\prime} = \sum_i \hat{A}_i \hat{\rho} \hat{A}_i^{\dagger} .
\end{equation}
The Kraus operators $\{\hat{A}_i \}$ here may take a variety of forms and are related to POVM elements by the decomposition $\hat{\pi}_i = \hat{A}_i^{\dagger} \hat{A}_i $, which is not a one-to-one mapping - e.g. the transformations $\hat{A}_i = \ket{0}\bra{0} $ and $ \hat{A}_i = ( \ket{0} \bra{0} + \ket{1} \bra{0}) / \sqrt{2} $ will both correspond to the same probability operator. Different Kraus operators represent different updates. It should be noted here that a more in-depth categorisation of measurement theory is possible, with some work distinguishing between three levels of description \cite{daviesopen, mathlang}. Firstly the levels of outcomes only, represented by POVMs. Further information can be added to this if the post-measurement state is included, typically by Kraus operators. This latter description (the set of maps between initial and final states) is known as a measurement model. A finer level of detail is possible and this includes practical information on how the measurement is performed, i.e. directly or through a probe. Maps of this kind are called instruments \cite{daviesopen, mathlang}. While both instruments and measurement models can be developed as consequences of what follows, this will not be the language used.
\par
In deriving Eq. \ref{eq:choi} as the most general formulation, three conditions are required in order that the transformation is physical \cite{croketransformations, kraus, choicomplete}. The map must preserve positivity as well as not increase the trace. It must also be linear: if $\hat{\rho}_1$ is transformed to $\hat{\rho}_1^{\prime}$ and $\hat{\rho}_2$ to $\hat{\rho}_2^{\prime}$, it follows that $p_1 \hat{\rho}_1 + p_2 \hat{\rho}_2$ becomes $p_1 \hat{\rho}_1^{\prime} + p_2 \hat{\rho}_2^{\prime}$. Finally, the stronger condition of complete positivity ensures that the same map acting on a subspace of a larger system ensures that the total density matrix stays positive. An alternate way to understand this is that a single system may be entangled with the rest of the universe; it should not be the case that acting locally on a system can create an unphysical density matrix on a global scale. 
\par
From the above channel, Eq. \ref{eq:choi}, the target measure can be found. A quantum state $\hat{\rho}$ is prepared by the procedure $s$. Following this, information is extracted by a process $x$, consisting of two measurements represented by the POVMs $\{ \hat{\pi}_i^{(1)} =  \hat{A}_i^{\dagger} \hat{A}_i \} $ and $\{ \hat{\pi}_j^{(2)} =  \hat{B}_j^{\dagger} \hat{B}_j \} $ respectively. The probability distribution associated with the first and second measurements, \textit{given that} a particular preparation and measurement strategy has been applied, is
\begin{equation} \label{eq:twomeasurementrule}
P(A_i , B_j | s, x) =  \Tr ( \hat{A}_i \hat{\rho} \hat{A}_i^{\dagger} \hat{B}_j^{\dagger} \hat{B}_j)  .
\end{equation}
As first pointed out in Kraus's original work on the subject \cite{kraus}, a logical prequisite is that the role of the first measurement can be considered two ways - either as part of the overall measurement process or as part of an extended preparation procedure, $s^{\prime}$ consisting of $s$ plus the action of $\hat{A}_i$. Due to the cyclic property of the trace operation, this appears in the above equation in that the Kraus operators may act either upon the initial density matrix or the second POVM element, resulting in a `two-outcome' operator $ \hat{\pi}_{ij}^{(1,2)} =  \hat{A}_i^{\dagger} \hat{\pi}_j^{(2)} \hat{A}_i $.
\par
The main result of this work is to verify the uniqueness of Eq. \ref{eq:twomeasurementrule} as a measure assigning a number $ 0 < p <1 $ for two POVMs which is consistent with the usual understanding of conditional probabilities. Due to the change of state during the first measurement, the connection between outcome and updated state mean that this is a non-trivial generalisation of the Busch-Gleason theorem. This is done in such a way as to actively include retrodiction and interdiction, in which intermediary outcomes can be studied given information about both later and earlier results.

\section{Liouville space single measurements} \label{single}

It is clear from the above discussion that the trace operation and its generalisations are the functionals of interest. Such functionals are inner products on an appropriately defined space and indeed inner products are the natural objects appearing in probabilistic theories \cite{hardy}, as any linear map from a vector space to real (or infact complex) numbers is an inner product \cite{ishamgroups}. Liouville space $\mathcal{L}$ is introduced in an appendix and is defined such that inner products on this space correspond to trace operations in Hilbert space, for which reason it is employed here. It is convenient to demonstrate the Busch-Gleason theorem in this space, as a stepping stone to the extended process which is the main focus of this work. It will be seen that the probability $ P(A_i | s,x)$ for a single measurement outcome is uniquely given as the inner product of two vectors: one associated with the particular outcome of measurement $A_i$; the other with prior information including any knowledge of past history of the measured system $s$ and the details of the measurement procedure, $x$. Measurement outcomes are defined as vectors in $\mathcal{L}$, in the sense that vectors in this space are isomorphic to operators, a subset of which are POVM elements.
\par
As in earlier work \cite{barnettetal}, the proof proceeds from just two postulates that concern the properties of a functional $\omega$: a map from measurement outcomes $A_i$ to real numbers. The first of these is a requirement for outcome independence:
\begin{equation} \label{eq:postulate1}
p(A_i | s,x) = N(s,x) \omega( \kett{A_i} ) .
\end{equation}
Aside from a normalisation factor $ N(s,x) $, this statement says that the map assigns a number to each possible probability operator in the set which has no dependence on how the set is completed. A physical example is a beam-splitter \cite{loudon}: further action on the reflected portion of the beam should not change the probability of transmission. Secondly, the functional $ \omega( \kett{A_i})  $ is assumed to be linear, so that 
\begin{equation} \label{eq:postulate2}
\alpha \omega( \kett{A_i} )  +  \beta \omega( \kett{B_j} )   =   \omega( \alpha \kett{A_i}  + \beta \kett{B_j}  ) ,
\end{equation}
with $\alpha$ and $\beta$ both non-negative numbers. In fact the effect of multiplying by these two constants need not be postulated; it may be derived from the case in which $\alpha=1=\beta$. However, this task is already performed in previous work \cite{busch, barnettetal} and need not be repeated here. 
\par
These postulates can both be justified by considering the number of \textit{counts} rather than probabilities, for an experiment which is rerun a finite number $N$ times. The measuring device will consist of a set of possible outcomes (for example, photodetector counts) denoted $i$, with each occuring $N(i)$ times. To go from this description to a probabilistic one, the set $\{N(i)\}$ is normalised by a factor of $1/N$, and the limit of large $N$ taken. From such a set up, the linearity condition follows from the fact that $ N(i\:{\rm OR}\:j) = N(i) + N({j}) $. The sense in which using vectors to describe measurements is natural (as in Hardy's work on axiomatic quantum mechanics \cite{hardy}) is that, if both $i$ and $j$ are allowed outcomes, then the combination $i {\rm OR} j$ is also allowed.
\par
The measurement is  decomposed into a set  $ \{\kett{a_j} \} $ of orthonormal basis vectors:
\begin{equation} \label{eq:decompose}
\kett{A_i} = \sum_j  \bbrakett{a_j}{A_i} \kett{a_j}, 
\end{equation}
which is acted upon by the required functional, Eq. \ref{eq:postulate2}, for
\begin{equation}
\omega( \kett{A_i} ) =   \sum_j \omega ( \kett{a_j} ) \bbrakett{a_j}{A_i} .
\end{equation}
The functional has been acquired and, as may be expected for assigning numbers to a vector space, is an inner product. Furthermore, another object $ \kett{r} = \sum_j \omega(\kett{a_j}) \kett{a_j} $ is that used in calculating these numbers. While at first glance this may seem fixed by the choice of basis, it should be noted that $\kett{a_j} =\kett{i k^{\dagger}} $ is in fact a two-coordinate vector such that there is an extra degree of freedom for the values $\omega(\kett{a_j}) $ in $\kett{r}$, which will be shown to be the analog in $\mathcal{L}$ to the density operator. This vector is normalised by summing over all probabilities $P(A_i | s,x) $. The vector $ \kett{A}  = \sum_i \kett{A_i}  $ is introduced, and the constant $ N(s,x) $ is found as
\begin{equation} \label{normalisation}
N(s,x) = \frac{1}{ \bbrakett{A}{r} }.
\end{equation}
Combining the above results, it has been shown that the probability associated with a single measurement is now a derived quantity: it is given by the inner product
\begin{equation} \label{quantum probability measure}
P(A_i | s,x) = \frac{ \bbrakett{A_i}{r} }{\bbrakett{A}{r} } 
\end{equation}
between the operator's Liouville space vector and a positive vector, identified as the density matrix vector $\kett{\rho} = \kett{r} / \bbrakett{A}{r} $. To clarify, it is only at this point that the notion of positivity of inner products enters. This is therefore the point at which quantum mechanics is introduced, by associating the vectors $\kett{A_i}$ with positive operators in Hilbert space. The allowed vectors $\kett{r}$ are all those for which $\bbrakett{A_i}{r} \geq 0$ for all measurements $\kett{A_i}$. If $\kett{A_i}$ correspond to positive operators on a Hilbert space then $\kett{r} $ also belongs to the set of positive operators. Explicitly, the probability rule is
\begin{equation}
P(A_i | s , x) =  \bbrakett{A_i}{\rho} .
\end{equation}
Its form depends on how the measurement is performed, as well as the probability of different outcomes, due to the preparation procedure. Both of those influences are characterised \textit{before} the measurement, hence $\kett{\rho} $ is understood as representing the prior information.  
\par
Earlier an identification between inner products in Liouville space and the trace operation in Hilbert space was made, shown in Eq. \ref{eq:liouvilletrace}. From this the above result can be shown to be directly equivalent to the usual, and expected, form for the probability rule. In this sense, the single measure probability rule has been derived rather than postulated; it is implicit within the Hilbert space structure.

\section{Main result: Conditional probabilities}

The probability distribution associated with a two-outcome process is provided by Kraus' rule, Eq. \ref{eq:twomeasurementrule}. In this section it is shown that this is the only possible probability measure given the constraint of complete positivity. For reasons discussed in the introduction, this uniqueness theorem places limits on fundamental and applied aspects of quantum physics. 
\par
As mentioned earlier, the analogous proof for projective measurements has already been supplied \cite{cassinelli+zanghi}. The method employed in that work is to posit a new functional which behaves as the original, single-measurement case (i.e. it must be the trace of some operator product) but is more restricted, in the sense that it must also correspond in some way to the familiar notion of conditional probabilities familiar from Bayes. This materialises in requiring the functional to act on the overlapping component of two operators $\hat{E}$ and $\hat{F}$, followed by a demonstration of uniqueness. Central to this is the fact that an orthogonal operator $\hat{F}^{\prime} = \hat{I} - \hat{F} $, associated with eigenvectors of zero overlap with those of $\hat{F}$, may be used to decompose $\hat{E}$. Importantly, this latter technique will not be permissible for an extension to generalised measurements, as POVM elements do not satisfy the property of non-overlapping eigenvectors. For this reason, a different approach is taken in what follows. Another motivation for taking a different pathway is to remove the implicit linear causality. Any approach which takes an initial state, updates it by a particular rule (i.e. the Choi map \cite{choicomplete}) and then acts on this assumes that the process may act only one way. In fact, this is known to be false. Retrodictive quantum mechanics swaps the traditional role of measurement and preparation, so that a known outcome may `evolve' back and undergo state collapse onto a mixed state of preparations. Kraus operators, however, still act in the same way. Further complexity occurs in what is referred to here as `interdiction'; knowledge about the initial and final states based on the first of two measurements. One might anticipate, and in fact it is found, that the Kraus formalism still holds.
\par
In what follows, the notation that will be used is that a preparation procedure giving density operator $\kett{\rho}$ is followed by two measurements with results denoted $A_i$ and $B_j$ respectively. The first measurement (with output $A_i$) will act as a transform; this is one of the key insights of quantum mechanics. As mentioned before, and originally argued by Kraus, this can be thought of as a transformation acting upon either the preparation procedure or second measurement. However, it does not follow directly from this that transform will be linear, so that it can be treated as a superoperator acting upon either measurement. Instead, this is justified by returning to the idea of vectors as representing count rates rather than probability distributions. In particular, consider a subset of outputs from the first measurement; the effect on the count rate for $B_0$ given postselection of two outcomes $A_0$ and $A_1$. This is patently linear in the sense that $N(B_0 , A_0 \: { \rm OR } \: A_1) = N(B_0, A_0) + N(B_0, A_1) $. Given Kraus's argument, the transformation is then viewed as acting upon the second measurement, which translates the previous sense of linearity into the possibility of associating a superoperator $ \hat{\hat{A}}_i $ with the first measurement. The overall probability rule must take the form
\begin{equation} \label{eq:liouvilleoverlap}
P(A_i, B_j | s , x) =  \bbra{ B_j}  \hat{\hat{A}}_i \kett{\rho} ,
\end{equation}
with the task being to constrain the superoperator such that it is consistent with the familiar notions of probability measures which were formalised in the previous section. As a first step it is convenient to explicitly map the above onto inner products in a different space. This can be established by noting that it may be re-expressed as a trace in Liouville space:
\begin{equation} \label{eq:transformationrule}
P(A_i, B_j | s , x) = \Tr_L (  \hat{\hat{A}}_i \kett{\rho} \bbra{ B_j} ).
\end{equation}
As Liouville space inner products correspond to the trace taken in Hilbert space, a higher dimensional `super-Liouville' space $\mathcal{S}$ can be invoked so that inner products are traces in Liouville space. The properties of this construct are discussed in an appendix, where it is formally defined, and vectors in the space are notated with double parentheses: $|\cdot))$. The probability rule is now 
\begin{equation} \label{eq:superliouvilleoverlap}
P(A_i, B_j | s , x) = (( A_i | B_j, \rho )).
\end{equation}
This identification is important, as it places a restriction on which vectors in the space may be identified as superoperator. In general, the transformation vector will take the form
\begin{equation} \label{eq:mostgeneral}
|A_i)) = \sum_{\alpha \beta k l } A^{\alpha \beta}_{k l} \kett{\alpha \beta^{\dagger}} \kett{ \tilde{k} \tilde{l}^{\dagger}} ,
\end{equation}
however not all choices for the set of coefficients $ A^{\alpha \beta}_{k l} $ will correspond to physical processes, by which it is meant that all inner products as shown in Eq. \ref{eq:superliouvilleoverlap} are required to be positive numbers. As mentioned above, there are two senses in which this must be true: the measured state considered in itself and the more stringent complete positivity; the state entangled with another system. How these conditions manifest is seen most clearly by deriving the form of $ | \rho, B_j )) = \kett{\rho} \kett{\tilde{B}_j} $ in both cases (note that this formulation suggests that $\mathcal{L}$ is the space of preparations and that $\tilde{\mathcal{L}}$).
\par
For simplicity a pure state will be considered here, such that the Liouville space density operator is $\kett{\rho} = \sum_{nm} c_n c_m \kett{n m^{\dagger}}$; if instead a mixed state is chosen then the two coefficients would be replaced by a single $c_{nm}$ but this does not change what follows. Using the same basis, the later measurement vector is written as $\kett{B_j} = \sum_{mn} B_m B^{*}_n \kett{m n^{\dagger}}$, where pure state projectors are again considered. Here, the set of $c_n$ can be chosen to be non-negative but in general $B_n$ will be complex. The vector $ | \rho, B_j )) $ is then 
\begin{equation} \label{eq:apriorivector}
| \rho, B_j )) = \sum_{m n p r} c_m c_n B_p^{*} B_r \ket{m} \ket{n^\dagger} \ket{\tilde{p}} \ket{\tilde{r}^{\dagger}}.
\end{equation}
It is interesting to make a comparison between the current construction and the `two time formalism' \cite{2svf}, which employs vectors of the form $\ket{\psi}_{t1} \otimes \bra{\phi}_{t2}$ associated with a single system at two different points in time $t_1$ and $t_2$. Such vectors also appear here, as they are quantities in the space $\mathcal{H}\otimes\mathcal{\tilde{H}}$ which is defined in an appendix. Importantly, the sense in which such vectors are employed differs in the two schemes: here, they are always in tensor product with the related vector on the dual space and no explicit physicial meaning is associated with the vectors in themselves. 
\par
The identity transformation $|I))$ can be naturally derived at this point, as an explicit example of a transformation in $\mathcal{S}$. If no measurement is performed between the preparation and second measurement, it may also be considered as a single measurement process. That is, the identity must be such that $ (( I | \rho, B_j)) = \bbrakett{\rho}{B_j} $ will hold for any $\kett{\rho}$ or $\kett{B_j}$, which implies that the identity takes the form $| I )) = \sum_{nm} |n \tilde{n} ) | m^{\dagger} \tilde{m}^{\dagger} ) $: it will factor into vectors on the product space $\mathcal{H}\otimes \tilde{\mathcal{H} }$. This will be seen to be a precursor to the full form that a transformation must take. In the specific case of a two-level system, this identity is
\begin{equation} \label{eq:id}
|I)) = |0 0^{\dagger} \tilde{0} \tilde{0}^{\dagger} )) + |0 1^{\dagger} \tilde{0} \tilde{1}^{\dagger} )) + |1 0^{\dagger} \tilde{1} \tilde{0}^{\dagger} )) + |1 1^{\dagger} \tilde{1} \tilde{1}^{\dagger} ))  .
\end{equation}

\begin{figure}
 \includegraphics[width=0.5\textwidth]{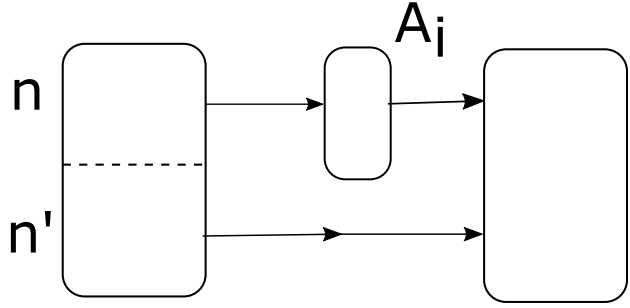}
 \caption{
 A general completely positive transform is visualised. The Schmidt decomposition of a vector in one space (represented by the solid black square) expresses it as the tensor product of two other spaces (according to the dashed line) with bases $\ket{n}$ and $\ket{n^{\prime}}$ respectively. Only the former space undergoes the transformation $A_i$. The condition of complete positivity enforces that the updated state (block on the left) must have a positive valued density matrix.} \label{fig:cptransform}
 \end{figure}
 
Eq. \ref{eq:apriorivector} is to be compared with the equivalent object which is derived from measuring only vectors on a subspace of the overall Hilbert space, shown in Fig. \ref{fig:cptransform}. A bipartite state in the Hilbert space may be decomposed into a Schmidt basis as $ \ket{\rho} = \sum_n c_n \ket{n}\ket{n^{\prime}} $, in which the dashed space is not acted upon. Using the same basis for the measurement eigenvector gives $\ket{B_j} = \sum_{mn} B_{mn} \ket{m}\ket{n^{\prime}} $. Again $c_n$ can be chosen to be positive but such a choice is not available for $B_{mn}$, which is complex in general. The vector which this is taken in inner product with will be $ | A_i \otimes I^{\prime} )) = |A_i)) \otimes |I^{\prime} )) $; the effect of the identity vector is to trace out the dashed subsystem: if the bipartite super-Liouville vector is denoted by an overline, then it is found that $ (( A_i, I^{\prime} | \overline{ \rho, B_j} )) = (( A_i | \rho, B_j)) $. In this object,
\begin{equation} 
| \rho, B_j )) = \sum_{m n p r} c_m c_n B_{pm} B_{rn}^{*} \ket{m} \ket{n^\dagger} \ket{\tilde{p}} \ket{\tilde{r}^{\dagger}}.
\end{equation}
It is seen that the additional requirement of complete positivity adds an extra index to the space; this has the result of requiring that allowed vectors must factorise into a vector $|\psi) = \sum_{mp} c_m B_{pm} |m) |\tilde{p}) $ on the doubled space $\mathcal{H} \otimes \tilde{\mathcal{H} }$. This is a different Liouville space, hence the different notation used, however there is still a mapping between Hilbert space operators and vectors. It is important, moreover, that any $|\psi)$ on the doubled space $\mathcal{H}\otimes \tilde{\mathcal{H}}$ is an allowed vector, including those which do not factor. These are analogous to entangled states and arise directly as a result of imposing complete positivity. For this reason, it can be stated with full generality that all physical transformations can be expressed in the current super-Liouville construction as
\begin{equation} \label{eq:transformvector}
|A_i)) = \sum_\chi \chi \sum_{\alpha k} \Lambda_{\alpha k}^{(\chi)} | \alpha \tilde{k} ) \sum_{\beta l} \Lambda_{\beta l}^{(\chi)*} | \beta^{\dagger} \tilde{l}^{\dagger} ),
\end{equation} 
in which the coefficients $ \{ A^{\alpha \beta}_{k l} \}$ from Eq. \ref{eq:mostgeneral} have been evaluated by spectral decomposition over the eigenvalues $ \{ \chi \} $. These correspond to positive operators on the space $\mathcal{H}\otimes\tilde{\mathcal{H}} $, and are analogous to the Choi-Jamiolkowski isomorphisms \cite{choicomplete, jamiolkowski} employed in quantum information theory, although they have entered here in a different way. The first sum corresponds to the Kraus operator from Eq. \ref{eq:choi} that multiplies from the left hand side and the second is that which multiplies from the right hand side.
\par
That vectors factorise in the way described above is the equivalent statement that one operator is the conjugate of the other. The probability rule is seen by explicitly calculating the inner product using equations \ref{eq:apriorivector} and \ref{eq:transformvector} to be 
\begin{equation}
(( A_i | \rho, B_j )) = \sum_{\chi} \chi \sum_{mnkl} \Lambda_{mk}^{(\chi)} \Lambda^{(\chi)*}_{nl} c_m c_n B_k B_l^{*}.
\end{equation}
Identifying $\rho_{mn} = c_m c_n$ and $B_{kl} = B_k B_l^{*}$ as the matrix elements of the relevant density matrix and second measurement POVM element respectively, this may be rewritten as 
\begin{equation}
 ((A_i | \rho, B_j)) = \sum_{\chi} \chi \sum_{mnkl} \Lambda^{(\chi)}_{m k} \rho_{mn} \Lambda^{(\chi)*}_{n l} B_{k l} = \sum_{\chi} \Tr( \hat{\Lambda}^{(\chi)} \hat{\rho} \hat{\Lambda}^{(\chi) \dagger} \hat{B}_j ) .
\end{equation}
 It is now possible to derive a Gleason-Busch type conditional probability rule for sequential measurements analagous to the single measurement rule derived in Section \ref{single}. So far it has been shown that joint probability of outcomes $A_i$ and $B_j$, given some preparation $s$ and context $x$, may be expressed as an inner product:
\begin{equation}
P(A_i, B_j | s, x) = (( A_i | \rho, B_j)).
\end{equation}
Furthermore, it has been shown that the constraint of positivity restricts those vectors $|A_i))$ that may represent completely positive transformations. In this way, the Choi-Jamiolkowski isomorphisms were derived, consistent with a small number of physically motivated first principles.  
\par
Ideally, one would like to logically separate in the inner product that which is related to the particular measurement outcome from that which describes all prior information (concerning state preparation, outcomes of intermediate or later measurements and measurement context). The context of measurement $A_i$ and $B_j$ is denoted by $x_A$ and $x_B$ respectively; this may include details of how the measurement is performed physically as well as the set of other possible outcomes. Bayesian reasoning requires that conditional and joint probabilities are related in such a way that all are proportional to the same inner product:
\begin{align}
P(A_i, B_j | s, x_A, x_B) \propto ((A_i | \rho, B_j)), \\
P(A_i | s, B_j, x_A) \propto ((A_i | \rho, B_j)), \\
P(B_j | s, A_i, x_B) \propto ((A_i | \rho, B_j)).
\end{align}
The relevant constants of proportionality will be different in each case and may be calculated using the requirement that probabilities sum to one. The normalised joint probability distribution is
\begin{equation} \label{eq:jointprob}
P(A_i, B_j | s, x_A, x_B) = \frac{ ((A_i | \rho, B_j)) }{\sum_{ij} ((A_i | \rho, B_j))}.
\end{equation}
This is the probability of obtaining outcome $A_i$ in the first measurement and $B_j$ in the second, conditioned on all prior information $s$, $x_A$ and $x_B$. Such a series of outcomes may also be considered as a single measurement procedure with outcomes indexed by both $i$ and $j$ by contracting vectors into a single Liouville space overlap, consistent with the idea of single measurements outlined in Section \ref{single}. That is, 
\begin{equation} \label{eq:jointprob2}
P(A_i, B_j | s, x_A, x_B) = \bbrakett{A_i B_j}{R}
\end{equation}
in which $\kett{A_i B_j} = \bbra{\tilde{B}_j} A_i ))$ and $ \kett{R} = \kett{\rho} / \sum_{ij} ((A_i | \rho, B_j)) $. The denominator in $ \kett{R}$ is simply that required for normalisation of the conditional probability in Eq. \ref{eq:jointprob2}.For an interdictive measurement, the probabilities summed over are on the first measurement only; the required inner product is
\begin{equation}
P(A_i | s, B_j, x_A ) = \frac{ ((A_i | \rho, B_j)) }{\sum_{k} ((A_k | \rho, B_j))} = ((A_i | R_I),
\end{equation}
with $|R_I)) = | \rho, B_j )) / \sum_{k} ((A_k | \rho, B_j))$, defined similarly to the previous case. Here the prior information is the pre- and post-selection on $\rho$ and $B_j$, along with the set of intermediate outcomes $ \{ A_k \} $. Note that these need not be a complete set, the probability rule is general enough to accomodate this.
\par
Finally, if the intermediate outcome is known, the conditional probability for the second measurement is obtained as
\begin{equation} \label{eq:vectorkraus}
P(B_j | s, A_i, x_B) = \frac{ ((A_i | \rho, B_j)) }{\sum_{k} ((A_i | \rho, B_k))} = \bbrakett {\tilde{B}_j}{\tilde{R}^{\prime}},
\end{equation}
with $ \kett{\tilde{R}^{\prime}} = (( A_i \kett{\rho} / \sum_k ((A_i| \rho, B_k)) $: this is the update rule corresponding to Eq. \ref{eq:choi}. 
\par

\section{Applications}
Until this point the focus has been a uniqueness theorem for the formalism of Kraus operators for sequential measurements. A new tool, the use of super-Liouville space to handle measurements, was introduced for this purpose. Such a construct is a natural way to handle interdictive measurements in particular and may be thought of in a Bayesian manner in the sense that the objects of interest are the probabilities for experimental outcomes given prior and also subsequent events. The relationship between such conditional probabilities is governed by Bayes' theorem. 
\par
Here a number of applications of the new formalism are demonstrated, examples of calculations performed in both Liouville and super-Liouville space. The operation of partial transposition in super-Liouville space is constructed, making clear the sense in which unphysical maps are disallowed. The BB84 protocol is analysed showing an axiomatic approach to quantum security: this is a case in which foundational issues become key to practical implementations.The quantum eraser provides an example of how to perform calculations using the new framework, particularly in cases for which the interdictive measurements are key. 

\subsection{Partial transposition}
To begin, an unphysical map is demonstrated, the classic example of which is a partial transpose. For a bipartite state the density matrix of one system only is transposed and it is found that this density matrix can be positive while creating a non-positive (and hence disallowed) density matrix for the whole system \cite{steveqi}. The transpose is a map that acts as follows:
\begin{align*}
\mathcal{T} ( \ket{0}\bra{0} )  \rightarrow \ket{0}\bra{0} \\
\mathcal{T} ( \ket{0}\bra{1} ) \rightarrow \ket{1} \bra{0} \\
\mathcal{T} ( \ket{1} \bra{0} ) \rightarrow \ket{0} \bra{1} \\
\mathcal{T} ( \ket{1} \bra{1} ) \rightarrow \ket{1} \bra{1} .
\end{align*}
For a two-qubit system, the transpose can be represented by a vector in super-Liouville space as
\begin{equation}
| T )) = | 0 0^{\dagger} \tilde{0} \tilde{0}^{\dagger} )) + | 0 1^{\dagger} \tilde{1} \tilde{0}^{\dagger} )) + | 1 0^{\dagger} \tilde{0} \tilde{1}^{\dagger} )) + | 1 1^{\dagger} \tilde{1} \tilde{1}^{\dagger} )) .
\end{equation}
Such an object cannot be expressed in the product form derived above and hence is not an allowed measurement procedure. This can be seen explicitly by considering a particular physical process: a state is prepared in the two-qubit Bell state $ \ket{\Phi^{+}} = ( \ket{0}_A \ket{0}_B + \ket{1}_A \ket{1}_B ) / \sqrt{2} $ and, after a partial transpose represented by the vector $ | T , I )) = | T ))_A |I))_B $ (so that the transpose only acts on qubit $A$) , a projective measurement is performed with outcome $ \ket{\Psi^{-}} = ( \ket{0}_A \ket{1}_B - \ket{1}_A \ket{0}_B ) / \sqrt{2} $.  The probability of this series of events will be given by the overlap
\begin{equation}
(( T, I | \Phi^{+}, \Psi^{-} )) = - \frac{1}{2}.
\end{equation}
Probabilities cannot be negative and hence the unphysical nature of partial transposition has been demonstrated.

\subsection{The BB84 Protocol}

The first and most well-known quantum cryptographic routine, due to Bennett and Brassard and commonly known as BB84 \cite{BB84}, has the aim of generating a key with which messages may be encrypted. Alice and Bob are two parties with access to a quantum and classical channel, the difference in function between the two being that eavesdroppers in the former can be detected while not in the latter. Alice encodes a string of logical bits $0$ or $1$ onto one of two choices of orthogonal states for a qubit, either $ \{ \ket{0}, \ket{1} \}$ or $\{ \ket{+} = (\ket{0} + \ket{1})/\sqrt{2} , \ket{-} = (\ket{0} - \ket{1})/\sqrt{2} ) \}$. One example of how these states could be physically implemented would be using a photon, with the two bases corresponding to vertical-horizontal versus the two diagonal polarisations. Bob is not aware of the basis chosen and so picks randomly between the two in which to perform a von Neumann measurement; in the case that his choice matches that of Alice then his measurement will give the correct value, assuming no noise or eavesdropper. Finally the two parties use the classical channel to mutually announce the basis used on each bit and in the cases with disagreement the bits are removed, leaving them with a binary string which is used as the cryptographic key. 
\par
The picture is complicated by the potential for an eavesdropper, Eve. This third party has the ability to detect both channels. Of particular interest is the quantum channel, for the only method available to extract information is to perform a quantum measurement, which will introduce the possibility for Bob and Alice to disagree on the bit value even in cases in which they both use the same basis. This is central to the function of the protocol, as it allows Alice and Bob to detect the prescence of Eve by announcing a sample of bits classically. This is made more difficult by the unavoidable prescence of noise on the channel, which will introduce a level of errors. Following this routine, Alice and Bob perform classical privacy amplification and error correction algorithms.
\par
Finding Eve's optimum measurement strategy is clearly an important task and has two aspects: extracting maximum information and minimising the quantum bit error rate which is introduced by her scheme. In this section the focus will be on the first of these. Using the Liouville space framework derived in the previous sections, the optimal strategy for Eve to determine the bit value sent by Alice will be calculated. The only limitation will be that her measurements are restricted to individual qubits. This situation has been studied many times with the resultant strategy well-known. Two results are derived in what follows. The probability $P(A=E=B)$ - i.e. that all three parties agree on the bit value - is maximised irregardless of Alice and Bob's results. Then the modified case that Eve has the extra knowledge of Alice and Bob's agreement is analysed, where it is shown that this leads to the previously discussed measurement. Though both of these strategies are already discussed in the literature, both appear naturally in the Liouville space formalism. 
\par
Analysis of the BB84 protocol is an example of a calculation which appears naturally as an eigenvalue problem in Liouville space. The joint probability rule Eq. \ref{eq:jointprob} is first re-written in such a manner. While it need not be the case for a general procedure, a further assumption is made that Eve's measurement is trace-preserving, which is expressed in the form
\begin{equation} \label{eq:tracepreserving}
\sum_i \bbra{I} A_i )) = \kett{\tilde{I}},
\end{equation}
a contraction of the vector from $\mathcal{L} \otimes \tilde{\mathcal{L}} $ into $ \mathcal{L}$. Physically, trace preservation conserves the sum of probabilities over the set of possible events, so that in this case $  \sum_{ij} (( A_i | \rho, B_j )) = 1 $. Eq. \ref{eq:jointprob} is now written as 
\begin{equation}
P(A_i, B_j | s, x) = \sum_\chi \chi ( A_i^{(\chi)} | \rho, \tilde{B}_j ) ( \rho^{\dagger}, \tilde{B}_j^{\dagger} | \tilde{A}_i^{(\chi)} ) ,
\end{equation}
with all vectors in this equation on the space $\mathcal{H} \otimes \tilde{\mathcal{H}} $. In this sense Eve's task is clearer: their best chance to avoid detection is to maximise the overlap of the Kraus operator vector with this preparation-measurement vector. 
\par
In the specific case of the BB84 protocol, it is known to Eve that Alice and Bob will discard all cases in which the preparation and second measurement do not match through a process of sifting and error correction. The object to consider is thus
\begin{align} \label{eq:opt1}
P(A=E=B) = \frac{1}{4} \sum_i (A_0^{i}| \big[ |0,\tilde{0} )(0^{\dagger} \tilde{0}^{\dagger}| + |+, \tilde{+} ) ( +^{\dagger}, \tilde{+}^{\dagger} | \big] | A_0^{i} )  \\
+  \frac{1}{4} \sum_i (A_1^{i}| \big[ |1, \tilde{1} ) (1^{\dagger}, \tilde{1}^{\dagger} | + |-, \tilde{-}) (-^{\dagger}, \tilde{-}^{\dagger} | \big] |A_1^{i}),
\end{align}
with the sums over $i$ allowing for the possibility that multiple measurement outcomes are associated with each bit value. The following analysis will make the assumption that there is only a single operator in each case, i.e. that the index takes the value $i = 1$ only in both cases, and so will be dropped.  
\par
Clearly the inner product in Eq. \ref{eq:opt1} would be maximised if $|A_0)$ and $|A_1)$ are the normalised eigenvectors of the Liouville space operators $ |0,\tilde{0} )(0^{\dagger} \tilde{0}^{\dagger}| + |+, \tilde{+} ) ( +^{\dagger}, \tilde{+}^{\dagger} |$ and $|1, \tilde{1} ) (1^{\dagger}, \tilde{1}^{\dagger} | + |-, \tilde{-}) (-^{\dagger}, \tilde{-}^{\dagger} | $ respectively:
\begin{align}
|A_0 )  = \frac{1}{\sqrt{3}} \big[ | 0, \tilde{0} ) \pm |+, \tilde{+} ) \big], \\
|A_1) = \frac{1}{\sqrt{3}} \big[ | 1 \tilde{1} ) \pm |- , \tilde{-} ) \big].
\end{align}
While both positive and negative choices in the above sums are eigenvectors, those which are positive are associated with the largest eigenvalues and so this choice gives the highest probability of success. It is found that these eigenvectors also satisfy trace-preservation condition and so are the optimal measurement; upon substitution the value $P(A=E=B)_{max} = 3/4$ is found. The other parameter characterising the measurement is the chance that Eve introduces disagreement between Alice and Bob's bit values --- it is by finding this value for a sample of their bits that it is possible for them to determine the existence (or not) of an eavesdropper. The probability that they do agree is the sum of the operators found above over each possible basis (represented by the summation index $i$):
\begin{equation}
P(A=B) = \frac{1}{4} \sum_i (i \tilde{i}| \big[ |A_0) (A_0| + | A_1)(A_1| \big] |i \tilde{i}) = \frac{5}{6}.
\end{equation}
The probability that these measurements do not match is hence $1/6$, and the fraction of the remaining bits that contribute will be $P(A=E=B | A=B) = 9/10$. Instead, Eve may be interested in maximising this latter value, so that (for the cases in which no errors are introduced) the correct bit value will be uncovered. It turns out that such a strategy exists that Eve gets the correct bit value in all such cases, with the trade-off that errors are introduced one-third of the time. The required probability is evaluated as 
\begin{equation}
P(A=E=B | A=B) = \frac{ (A_0| \big[ |0)(0| + |+)(+| \big] |A_0) + (A_1| \big[ |1)(1| + |-)(-| \big] |A_1)}{\sum_{i} (i| \big[ |A_0)(A_0| + |A_1)(A_1| \big]|i)},
\end{equation}
which may be simplified by noting that an exchange of index, from $0$ to either $1$ or $+$ or $1$ to either $0$ or $-$, should not alter any probabilities. Taking this into account results in the simpler form
\begin{equation}
P(A=E=B|A=B) = \frac{  ( 0 \tilde{0} | A_0) (A_0| 0 \tilde{0}) }{(0 \tilde{0}| \big[ |A_0)(A_0 | + |A_1)(A_1| \big] |0 \tilde{0}) }.
\end{equation}
It is seen that there will be complete agreement between all three parties if $ (0 \tilde{0}| A_1 )(A_1|0 \tilde{0} ) = 0$, such that the vector $|A_1)$ contains no $|0 \tilde{0})$ terms when written in the same basis as the preparation-measurement vector. There is only one such vector which also satisfies both the normalisation condition and the symmery detailed above (i.e. that labels $0$ and $+$ may be interchanged) and this is
\begin{equation}
|A_1) = \frac{1}{\sqrt{6}} \big[ |0 \tilde{1} ) + |1 \tilde{0} ) - 2 | 1 \tilde{1}) \big].
\end{equation}
By the same argument the measurement with outcome assigned to the bit value zero will be 
\begin{equation}
|A_0) = \frac{1}{\sqrt{6}} \big[ |0 \tilde{1} ) + |1 \tilde{0}) - 2 |0 \tilde{0}) \big],
\end{equation}
and the combination of the two corresponds to a measurement which introduces disagreement between Bob and Alice with probability $1/3$ but provides agreement between all three parties in all other cases. Such a strategy is the so-called Fuchs-Peres-Brandt attack: it corresponds to Eve entangling the Alice-Bob qubit with a controlled-NOT gate which acts in the Breidtbart basis \cite{kimetal, fuchsetal, fuchsperes, brandt} $ \{ \ket{0_B} = \cos(\pi / 8) \ket{0} + \sin(\pi/8) \ket{1} , \ket{1_B} = \cos(\pi/8) \ket{1} - \sin(\pi/8) \ket{0} \} $. Given that Alice and Bob agree on the bit value - which will occur in two-thirds of cases - then Eve will be able to determine the bit value with certainty.

\subsection{Quantum eraser}

\begin{figure}
\includegraphics[width=0.5\textwidth]{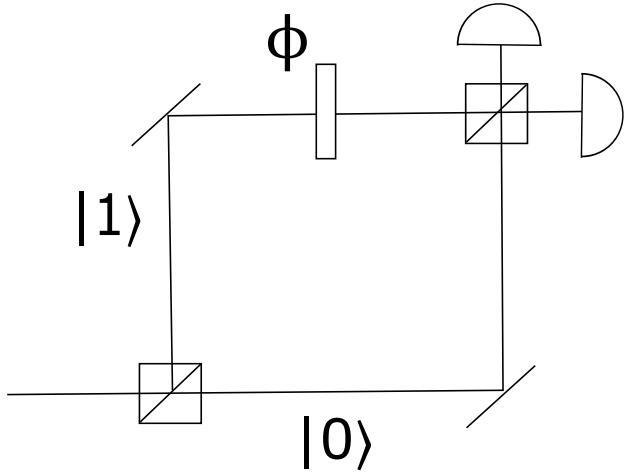}
\caption{An interferometer separates an input state $\ket{0}$ into two arms, along one of which a phase change of $\phi$ is introduced so that the new state is $\ket{0} +e^{i\phi} \ket{1}$. The two arms are then brought together through a final beamsplitter, after which detectors register counts. By changing the action of the later beamsplitter, this state being measured for is chosen.} \label{fig:machzehnder}
\end{figure}

The delayed choice quantum eraser is an experiment often presented as a prime example of quantum `weirdness' \cite{quantumeraser}. A photon enters an interferometer, as in Fig. \ref{fig:machzehnder}, before arriving at a screen and displaying an interference pattern. If the photon is initially prepared in a mode corresponding to the state $ \ket{0} $, the initial beamsplitter changes the state to $\ket{\rho} = (\ket{0} + e^{{i\phi}}  \ket{1} ) / \sqrt{2} $ (with a phase change of $\phi$ introduced by the difference in path-length between the two arms). A measurement of the path degree of freedom (i.e. in the basis $\ket{0}$, $\ket{1}$) gives each outcome with equal probability, independent of the phase $\phi$. If however the paths are recombined at a second beamsplitter before detection, equivalent to a measurement in the conjugate basis ($\frac{1}{\sqrt2}(\ket{0} \pm \ket{1} )$), the probabilities ${\rm P}(\pm) = \frac{1}{2} \left( 1 \pm \cos \phi \right)$ show an interference pattern, dependent on $\phi$. This simple set-up has been used to explore, both conceptually and experimentally, Bohr's principle of complementarity \cite{comperaser}: if information about the path degree of freedom is available, even in principle, information about the complementary basis, and therefore interference, is destroyed.

In delayed choice experiments, the choice of basis is made while the system is in flight, so that the information about whether the experiment should reveal the wave or particle-like nature is undetermined until the system to be measured is already in the experimental apparatus. The additional element in the quantum eraser is to show that if path information can be erased after the photons have been recorded, then an inteference pattern may be recovered. We use this example to demonstrate the use of super-Liouville space vectors to update probabilities given new information about what measurement was performed.

A device introduced before collapse to determine which path the photon travelled down may be modelled as an entangling qubit, which takes the probe-photon system into a Bell state $ (\ket{00} + e^{i \phi} \ket{11}) / \sqrt{2} $. The probe is thus perfectly correlated with the path degree of freedom and the ability to determine, in principle, which path the photon took destroys the interference. The actual measurement of the probe to determine the path taken can be conducted at any subsequent point, including long after the photon has been measured, or indeed not at all, with decoherence still occuring. If, on the other hand, the probe is measured in the conjugate basis, the interference pattern may be recovered, regardless of when the conjugate basis measurement is performed. While this may appear to allow for `backwards causality', the interference fringes left by the process require post-selecting a subset of all outcomes: this cannot be performed until after both measurements are done and thus causality is saved.

\par

Before introducing the delayed choice aspect, the interferometer analysis can be performed in the formalism of the new space. The density matrix is written as a Liouville space construction
\begin{equation}
\kett{\rho} = \frac{1}{2} \big[ \kett{0 0^{\dagger}} + e^{-i\phi} \kett{0 1^{\dagger}} + e^{i \phi} \kett{1 0^{\dagger}} + \kett{ 1 1^{\dagger}} \big]
\end{equation}
and, following the method outlined in the earlier parts of this paper, this is taken in product with another vector which is embedded in the complementary space $\tilde{\mathcal{L}}$ representing the POVM element corresponding to the later measurement outcome. For simplicity, here a projective measurement in the $\{ \ket{+} = ( \ket{0} + \ket{1} ) / \sqrt{2} , \ket{-} = ( \ket{0} - \ket{1} ) / \sqrt{2} \}$ basis will be used. The projectors describing this measurement correspond to vectors in $\tilde{\mathcal{L}}$:
\begin{align}
\kett{\tilde{+}} &= \frac{1}{2} \big[ \kett{ \tilde{0} \tilde{0}^{\dagger}} + \kett{ \tilde{0} \tilde{1}^{\dagger}} + \kett{ \tilde{1} \tilde{0}^{\dagger}} + \kett{ \tilde{1} \tilde{1}^{\dagger}} \big], \nonumber \\
\kett{\tilde{-}} &= \frac{1}{2} \big[ \kett{ \tilde{0} \tilde{0}^{\dagger}} - \kett{ \tilde{0} \tilde{1}^{\dagger}} - \kett{ \tilde{1} \tilde{0}^{\dagger}} + \kett{ \tilde{1} \tilde{1}^{\dagger}} \big],
\end{align}
so that the preparation-measurement vector is a tensor product of $\kett{\rho}$ with one of these two vectors. Two possibilities are now considered for the transformation which occurs between those two events: either there is no intervention, or an entangling probe learns about the path degree of freedom. In the first case, the overlap with the identity vector, Eq. \ref{eq:id}, is $ (( I | \rho, \tilde{+} )) = (1 + \cos{\phi} ) / 2 $ corresponding to the expected interference fringes. In the second case, our system is entangled with a probe, which is later measured. It is illustrative of the new method to translate the usual picture into the super-Liouville space formalism. In the usual quantum mechanical picture, the probe system learns about the path degree of freedom through a controlled-NOT operation, $\hat{U} = \ket{0} \bra{0} \otimes \hat{I} + \ket{1} \bra{1} \otimes \hat{\sigma}_x$. Thus the initial state $\frac{1}{\sqrt2} \left( \ket{0} + e^{i \phi} \ket{1} \right) \ket{0}$ evolves to $\frac{1}{\sqrt2} \left( \ket{0} \ket{0} + e^{i \phi} \ket{1} \ket{1} \right)$, the Bell state given earlier. Clearly tracing over the probe system destroys any coherence in the original photon, transforming a superposition of $\ket{0}$, $\ket{1}$ to a mixture, and may be described by a transformation which acts as follows:
\begin{align}
\mathcal{E} \left( \ket{0} \bra{0} \right) &= \ket{0} \bra{0} \\
\mathcal{E} \left( \ket{0} \bra{1} \right) &= 0 \\
\mathcal{E} \left( \ket{1} \bra{0} \right) &= 0 \\
\mathcal{E} \left( \ket{1} \bra{1} \right) &= \ket{1} \bra{1} .
\end{align}
In super-Liouville space this corresponds to the vector
\begin{equation}
|\mathcal{E})) = \ket{0} \ket{0^{\dagger}} \ket{\tilde{0}} \ket{\tilde{0}^{\dagger}} + \ket{1} \ket{1^{\dagger}} \ket{\tilde{1}} \ket{\tilde{1}^{\dagger}} .
\end{equation}
The overlap of the preparation and measurement vectors with this vector is $ (( \mathcal{E} | \rho, \tilde{+} )) = 1/ 2 $, and the interference pattern has disappeared, as expected. The erasure part of the protocol is now examined. If the probe system is later measured in some arbitrary basis $\ket{v_0}$, $\ket{v_1}$ the effect on the system alone is described via Kraus operators:
\begin{align}
\hat{A}_0 &= \bra{v_0} \hat{U} \ket{0} = v_{00} \ket{0} \bra{0} + v_{01} \ket{1} \bra{1}  \\
\hat{A}_1 &= \bra{v_1} \hat{U} \ket{0} = v_{10} \ket{0} \bra{0} + v_{11} \ket{1} \bra{1},
\end{align}
corresponding to super-Liouville vectors of the form
\begin{equation}
| A_i )) = \big[ v_{i0} |0 \tilde{0} ) + v_{i1} | 1 \tilde{1} ) \big] \otimes \big[ v_{i0}^* |0^{\dagger} \tilde{0}^{\dagger} ) + v_{i1}^* | 1^{\dagger} \tilde{1}^{\dagger} ) \big].
\end{equation}
where $v_{ij} = \braket{v_i}{j}$ are the elements of a unitary matrix. It is readily verified that $| \mathcal{E} )) = | A_0 )) + | A_1 ))$, as expected. All detection events may now be filtered into those for which the probe measurement gave outcome $A_0$ and those corresponding to outcome $A_1$. In particular, if the probe is measured in the $\ket{+}$, $\ket{-}$ basis, we obtain
\begin{align}
| A_+ )) &= \frac{1}{2} \big[ |0 \tilde{0} )) + | 1 \tilde{1} )) \big] \otimes \big[ |0^{\dagger} \tilde{0}^{\dagger} )) + | 1^{\dagger} \tilde{1}^{\dagger} )) \big] \\
| A_- )) &= \frac{1}{2} \big[ |0 \tilde{0} )) - | 1 \tilde{1} )) \big] \otimes \big[ |0^{\dagger} \tilde{0}^{\dagger} )) - | 1^{\dagger} \tilde{1}^{\dagger} )) \big] .
\end{align}
The first is proportional to the identity super-operator, reflecting the fact that path information is erased and coherence is restored. The second has only a phase difference between components compared to the identity, and up to a relabelling of outcomes also enables the recovery of an interference pattern. The conditional probabilities of interest, taken from Eq. \ref{eq:vectorkraus}, are:
\begin{align*}
P(+|A_+,\rho) &= \frac{(( A_+ | \rho, \tilde{+} ))}{(( A_+ | \rho, \tilde{+} ))+(( A_+ | \rho, \tilde{-} ))} \\
& = \frac{1}{2}(1+ \cos \phi) \\
P(-|A_+,\rho) & = \frac{1}{2}(1- \cos \phi) \\
P(+|A_-,\rho) &= \frac{(( A_- | \rho, \tilde{+} ))}{(( A_- | \rho, \tilde{+} ))+(( A_+ | \rho, \tilde{-} ))} \\
& = \frac{1}{2}(1- \cos \phi) \\
P(-|A_-,\rho) & = \frac{1}{2}(1+ \cos \phi)
\end{align*}
It is noted that the formalism is causally neutral: we can record the initial and final states first and later learn of the intermediate measurement, indeed in this case the intermediate measurement may be performed \emph{after} the detection of the photons, and the probability rule is equally valid.

\section{Conclusion}

The Kraus formalism has long been accepted as the method for handling sequential measurements and here it is demonstrated that, given some minimal assumptions, no other way is possible. There are a number of reasons that an axiomatic approach is important. These range from the foundations of the subject to more pratical issues of concern in quantum communications. The principal focus has been given to the consequences for secure communications: parties exchanging information encoded in qubits need to be sure that an alternative framework, so that intercepted messages could be altered in an unpredictable way, is not available. The axiomatic approach to deriving the properties of sequential measurements has the additional benefit of identifying those few axioms on which the security ultimately depends. This may be helpful in reassuring sceptical users of the new technology.

\appendix

\section*{Appendix: Complex vector spaces: a reminder}

Hilbert space is the stage upon which quantum mechanics plays out though this is, to some extent, a choice. Two other spaces can be used which highlight different aspects of the theory and will be employed throughout this work. As a reminder, here the mathematical background is covered.
\par
Any vector space is defined by a set of basis vectors $ \{ \ket{i} \} $, such that the `space' consists of any linear combinations of these vectors: $\{ \sum_i a_i \ket{i} \}$. Hilbert space is defined such that the coefficients $a_i$ are limited to be complex numbers. States of a system are associated with these vectors; operators (that is, maps between vectors) are associated with observable quantities. Restrictions upon these operators form the basic postulates of the theory. Importantly for the discussion here linear combinations also form operators; ergo, they may be represented in a vector space.
\par
 Liouville space $\mathcal{L}$ is defined as the tensor product of Hilbert space $\mathcal{H}$ and its dual \cite{liouville1, liouville2}. The two available indices allow for an isomorphism between the ket and bra of the operator and vectors in $\mathcal{L}$. A Hilbert space operator $\hat{A}$ is assigned a vector $\ket{A}\rangle$ in Liouville space, with direct correspondence
\begin{equation} \label{eq:liouvillehilbert}
\ket{i}\bra{j} \leftrightarrow \kett{ij^{\dagger}}  .
\end{equation}
As in the Heisenberg picture, system evolution here is associated with the operation rather than the state. The required mathematical object is a superoperator $\hat{\hat{L}}$ which acts upon Liouville space vectors, having the same physical effect as a map in $\mathcal{H}$.
 \par
In Gleason's theorem and its generalisations, the functional of interest is the trace: maps of the form $ \hat{A} \rightarrow \Tr ( \hat{A} \hat{B}) $. This is also the central motivation for invoking the alternatives to Hilbert space here. It is easily shown that scalar products in $\mathcal{L} $ are of the form
  \begin{equation} \label{eq:liouvilletrace}
  \bbrakett{A}{B} = \Tr_{ \mathcal{H} } ( \hat{A}^{\dagger} \hat{B} ),
 \end{equation}
where the subscript denotes a trace taken over the Hilbert space basis. A natural route is provided into the work by this fact --- through investigating ways in which the inner product can be preserved while generalising the overall object. It is seen on that this manifests into constraints upon superoperators which are placed inside the inner product.
\par
The third and final space utilised is a higher dimensional space referred to either as super-Liouville or transformation space, defined as the tensor product of a Liouville space with its dual space: $ \mathcal{S} = \mathcal{L} \otimes \mathcal{L}^{\dagger} = \mathcal{H} \otimes \mathcal{H}^{\dagger} \otimes \tilde{\mathcal{H}} \otimes \tilde{\mathcal{H}^{\dagger}}$ . To differentiate vectors in this space from those in $\mathcal{L}$, the `parenthetical ket' notation $ |x )) $ is used. For the same reason that an inner product in Liouville space corresponds to the Hilbert space trace, inner products in the new space are given by the trace operation in Liouville space. Vectors here are isomorphic to maps between the operators that act on Hilbert space vectors, though care should be taken with regards to using this space: not all vectors will correspond to physical transformation. \textit{Only a subset of all possible vectors, taken as inner products, can be interpreted as transformations.} Understanding this point will ultimately be what leads to the generality of Kraus operations.
\par
A physical example may help to understand the varied roles played by each space. A noisy measurement is used to distinguish between the two levels of a generic qubit, represented in Hilbert space by a vector $ \ket{\psi} = \alpha \ket{0} + \beta \ket{1} $, with a probability $p$ of measuring the wrong state; a scheme corresponding to the two-element POVM $ \{ \hat{\pi}_0 = (1-p) \ket{0}\bra{0} + p \ket{1}\bra{1} , \hat{\pi}_1 = p \ket{0}\bra{0} + (1-p) \ket{1}\bra{1} \} $. Considering just the first of these operators, that which means the zero outcome was detected, it is seen that the mapping from Eq. \ref{eq:liouvillehilbert} gives a Liouville space vector $ \kett{\pi_0} = (1-p) \kett{0 0^{\dagger}} + p \kett{1 1^{\dagger}} $. Following the measurement, the state will be updated with a degree of freedom in associating the update rule to the measurement outcome. There are three ways in which the update channel can be expressed. Firstly, as a Kraus operator, one choice of which is $ \hat{A}_0 = \sqrt{1-p} \ket{0}\bra{0} + \sqrt{p} \ket{1}\bra{1} $. This satisfies the relation $\hat{A}_0 \hat{A}_0^{\dagger} = \hat{\pi}_0 $ and corresponds to a post-measurement state of $\hat{A}_0 \ket{\psi} $ given that outcome. Secondly, as a superoperation $\hat{\hat{A}}_0 = \hat{A}_0 \otimes \hat{A}_0^{\dagger} $ which acts upon the Liouville space vector 
\begin{equation}
\kett{\psi} =  \ket{\psi} \ket{\psi^{\dagger}}  = |\alpha|^2 \kett{0 0^{\dagger}} + \alpha \beta^* \kett{0 1^{\dagger}} + \alpha^* \beta \kett{1 0^{\dagger}} + |\beta|^2 \kett{1 1^{\dagger}}
\end{equation}
of the density matrix. Thirdly, it will be shown that it may also be repesented as a super-Liouville space vector $| A_0 )) = |A_0) |\tilde{A}_0) $ (with the vectors $ |\cdot) $ vectors on the space $ \mathcal{H}\otimes \tilde{\mathcal{H}} $. This formulation does not act upon any state in the same way as the previous states; however, it will be seen later that taking products of this with specific objects in the space does have a meaningful interpretation in terms of conditional probabilities.

\section*{Acknowledgements}
This work was supported by the UK Engineering and Physical Sciences Research Council and by the Royal Society (RP150122) . We thank Mark Hillery for helpful comments and suggestions.

\bibliographystyle{apsrev4-1}
%

\end{document}